\documentclass[pra,amsfonts,amsmath,twocolumn,superscriptaddress,a4paper]{revtex4}

\usepackage[T1]{fontenc} \usepackage{ae} \usepackage{color} %\usepackage{times,mathptm}
\usepackage{graphicx}

\begin{document}

\title{Imaging electric fields in the vicinity of cryogenic surfaces using Rydberg atoms}

\author{T. Thiele} \email{tthiele@phys.ethz.ch} \affiliation{Department of Physics, ETH Z\"urich, CH-8093 Z\"urich, Switzerland}

\author{J. Deiglmayr} \affiliation{Laboratorium f\"ur Physikalische Chemie, ETH Z\"urich, CH-8093, Z\"urich, Switzerland}

\author{M. Stammeier} \affiliation{Department of Physics, ETH Z\"urich, CH-8093 Z\"urich, Switzerland}

\author{J.-A. Agner} \affiliation{Laboratorium f\"ur Physikalische Chemie, ETH Z\"urich, CH-8093, Z\"urich, Switzerland}

\author{H. Schmutz} \affiliation{Laboratorium f\"ur Physikalische Chemie, ETH Z\"urich, CH-8093, Z\"urich, Switzerland}

\author{F. Merkt} \affiliation{Laboratorium f\"ur Physikalische Chemie, ETH Z\"urich, CH-8093, Z\"urich, Switzerland}

\author{A.~Wallraff} \affiliation{Department of Physics, ETH Z\"urich, CH-8093 Z\"urich, Switzerland}

\renewcommand{\i}{{\mathrm i}} \def\1{\mathchoice{\rm 1\mskip-4.2mu l}{\rm 1\mskip-4.2mu l}{\rm
1\mskip-4.6mu l}{\rm 1\mskip-5.2mu l}} \newcommand{\ket}[1]{|#1\rangle} \newcommand{\bra}[1]{\langle
#1|} \newcommand{\braket}[2]{\langle #1|#2\rangle} \newcommand{\ketbra}[2]{|#1\rangle\langle#2|}
\newcommand{\opelem}[3]{\langle #1|#2|#3\rangle} \newcommand{\projection}[1]{|#1\rangle\langle#1|}
\newcommand{\scalar}[1]{\langle #1|#1\rangle} \newcommand{\op}[1]{\hat{#1}}
\newcommand{\vect}[1]{\boldsymbol{#1}} \newcommand{\id}{\text{id}}

\begin{abstract}
The ability to characterize static and time-dependent electric fields in situ is an important prerequisite for quantum-optics experiments with atoms close to surfaces. Especially in experiments which aim at coupling Rydberg atoms to the near field of superconducting circuits, the identification and subsequent elimination of sources of stray fields is crucial. We present a technique that allows the determination of stray-electric-field distributions $(F^\text{str}_\text{x}(\vec{r}),F^\text{str}_\text{y}(\vec{r}),F^\text{str}_\text{z}(\vec{r}))$ at distances of less than $2~\text{mm}$ from (cryogenic) surfaces using coherent Rydberg-Stark spectroscopy in a pulsed supersonic beam of metastable $1\text{s}^12\text{s}^1~{}^{1}S_{0}$ helium atoms. We demonstrate the capabilities of this technique by characterizing the electric stray field emanating from a structured superconducting surface. Exploiting coherent population transfer with microwave radiation from a coplanar waveguide, the same technique allows the characterization of the microwave-field distribution above the surface.
\end{abstract} 		\maketitle %\today \section{Introduction}

Hybrid systems aiming at coupling the internal state of atoms to solid-state devices have attracted significant interest in recent years. Realizations of such systems include neutral atoms close to atom chips~\cite{Fortagh1998,Vuletic1998,Drndic1998,Fortagh2002,Kruger2007,Boehi2010,Reichel2011,Boehi2012}, tapered fibers~\cite{Vetsch2010,Goban2012} or photonic waveguides~\cite{Tiecke2014,Goban2014}, Rydberg atoms close to mesoscopic devices~\cite{Sorensen2004,Petrosyan2008,Petrosyan2009} and ions near surfaces~\cite{Seidelin2006,Ospelkaus2008}. An important motivation for the development of hybrid systems is the combination of the long coherence times characteristic of atomic ensembles and the strong interactions and fast processing capabilities of solid-state devices as a route towards scalable quantum computing~\cite{Petrosyan2008,Petrosyan2009,Kurizkia2015}. Atomic and solid-state systems can be coupled by electromagnetic fields which have to be controlled with high accuracy. For example, in our experiment we aim at realizing strong coupling between Rydberg atoms and solid-state circuit QED devices~\cite{Wallraff2004,Chiorescu2004} using microwave fields.
Stray fields emanating from the patterned surfaces of solid-state devices in general are of major concern in hybrid systems involving ions~\cite{Seidelin2006,Ospelkaus2008} and Rydberg atoms~\cite{Sorensen2004,Petrosyan2008,Petrosyan2009}. The sources of stray fields are manifold and include surface adsorbates~\cite{Tauschinsky2010,Hattermann2012,Chan2014}, polycristalline surface patches~\cite{Carter2011}, and charges in the isolating gaps of coplanar waveguides (CPW)~\cite{Carter2012}. Although detrimental effects of electric stray fields might be mitigated in some cases, e.g., by microwave frequency dressing~\cite{Jones2013}, coating the surfaces with adsorbates~\cite{Hermann-Avigliano2014} or choosing chemically inert atoms (such as helium)~\cite{Hogan2012,Thiele2014}, techniques to measure stray fields are essential. A possible technique in this context is based on Rydberg-electromagnetically-induced transparency, with which also the microwave field in a glass cell~\cite{Holloway2014} or of a coplanar waveguide (CPW)~\cite{Sedlacek2012,Fan2014} has been characterized.

In this article, we present a technique to measure static and time-dependent (microwave) electric fields above patterned surfaces in a cryogenic environment based on coherent Rydberg-Stark spectroscopy. We use a fast ($v\approx 1700~\text{m}/\text{s}$) supersonic beam of metastable $1\text{s}^12\text{s}^1~{}^{1}S_{0}$ helium atoms. The atoms are excited to the $34\text{s}$ Rydberg state in two sequential, resonant one-photon transitions using a $10$-ns-long laser pulse of wavelength $\lambda\approx313~\text{nm}$ followed by a $160$-ns-long microwave pulse of frequency $\nu\approx27.966~\text{GHz}$~\cite{Thiele2014}. Specifically, we show that measuring spatially varying quadratic Stark shifts
\begin{equation}\label{eq:Stark}
\begin{aligned}
\Delta E_{\text{Stark}}^{(i)}(\vec{r})&=h(\nu_\text{a}(\vec{r})-\nu_0)= \frac{1}{2}\Delta\alpha |\vec{F}^{(i)}_\text{tot}(\vec{r})|^2\\
&=\frac{1}{2}\Delta\alpha |\vec{F}^{(i)}(\vec{r})+\vec{F}^\text{str}(\vec{r})|^2
\end{aligned}
\end{equation}
of the (conveniently chosen) $34\text{s}$ to $34\text{p}$ Rydberg-Rydberg transition for different well-defined applied electric fields [$(F^{(i)}_\text{x}(\vec{r}),F^{(i)}_\text{y}(\vec{r}),F^{(i)}_\text{z}(\vec{r}))$, $i=1-4$] allows the determination of the three components of an unknown stray field $(F^\text{str}_\text{x}(\vec{r}),F^\text{str}_\text{y}(\vec{r}),F^\text{str}_\text{z}(\vec{r}))$ emanating from the surface. Here, $\Delta \alpha=1078.03~\text{MHz}~(\text{V}/\text{cm})^{-2}$~\cite{Zimmerman1979,Thiele2014} is the polarizability difference between the $34\text{p}$ and $34\text{s}$ states. The stray field $\vec{F}^\text{str}(\vec{r})$ is uniquely determined at each point $\vec{r}$ by a set of four independent Equations [Eq.~(\ref{eq:Stark}), $i=1-4$]. Three equations are needed because of the three dimensions of the problem and an additional equation because of the quadratic nature of Eq.~(\ref{eq:Stark}). The spatially varying Stark shift $\Delta E_{\text{Stark}}^{(i)}(\vec{r})$ is observed as the difference between the position-dependent resonance frequency $\nu_\text{a}(\vec{r})$ of the $34\text{s}$ to $34\text{p}$ transition and its zero-field transition frequency $\nu_0=27\,965.773~\text{MHz}$, obtained from the known quantum defects~\cite{Drake1999}.

\begin{figure}[t] \centering \includegraphics[width=86mm]{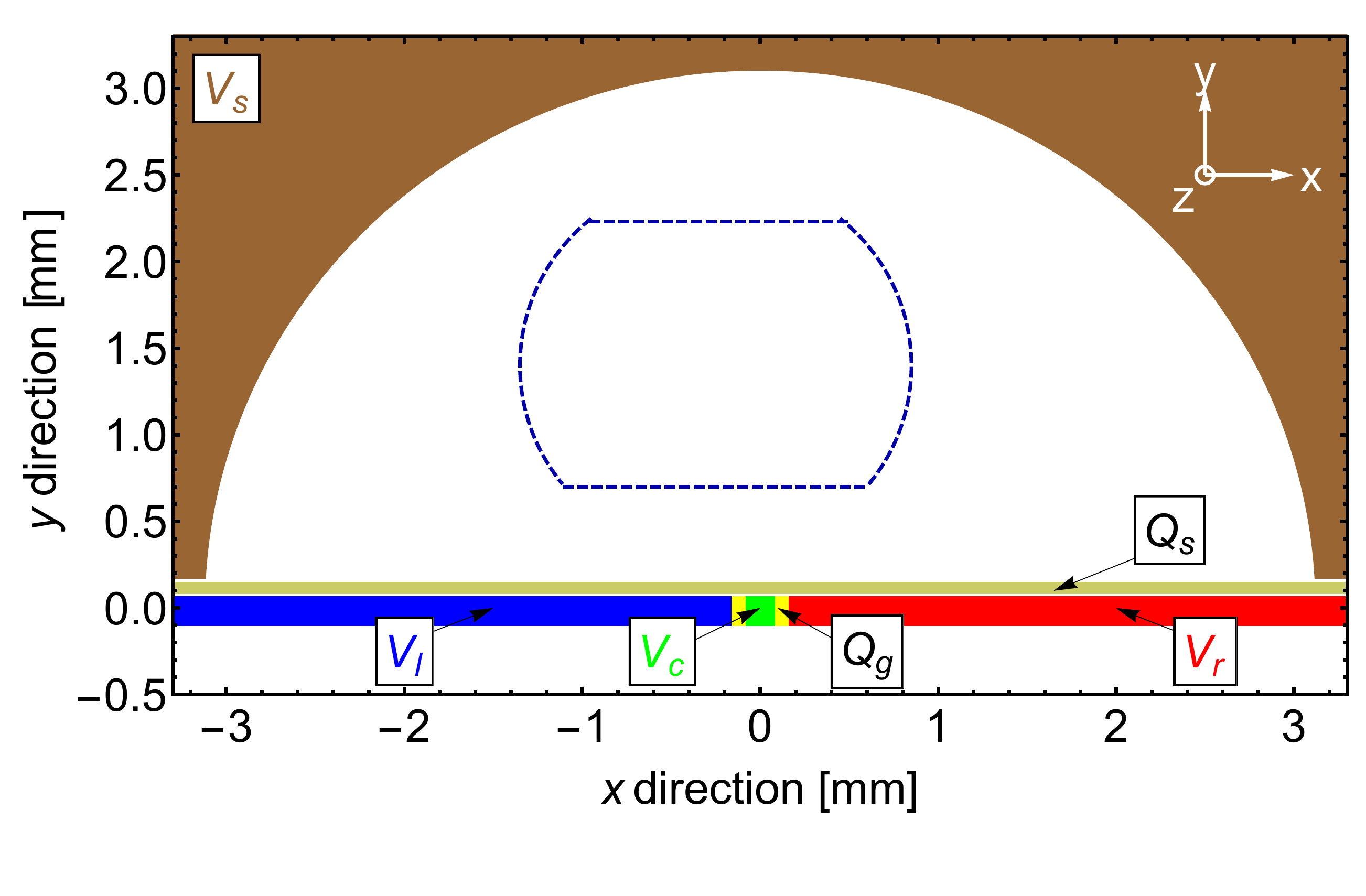} 	
\caption{Cross section of the interaction region above the coplanar waveguide. Potentials $V_\text{c}$, $V_\text{r}$, $V_\text{l}$ and $V_\text{s}$ can be applied to the center conductor (green), the right (red) and the left (blue) ground plane of the coplanar waveguide (CPW) and to the metallic shield (brown). The thickness of the superconducting layer of the CPW is not to scale. Charges $Q_\text{g}$ and $Q_\text{s}$ (light and dark yellow) may accumulate either in the insulating gaps of the waveguide or on an insulating layer on the superconductor. The dashed, blue line indicates the cross section of the Rydberg beam limited by the collimating slit apertures.}
\label{fig:Setup} \end{figure}

We perform the measurements in an experimental region cooled to $\sim 4~\text{K}$ that is translationally invariant over $6~\text{mm}$ along the helium-beam propagation direction ($z$ direction) and bounded in the $(x,y)$ plane by the surface of a patterned, $150$-$\text{nm}$-thick superconducting NbTiN film on sapphire and a metallic shield (Fig.~\ref{fig:Setup}). The superconducting film contains a coplanar waveguide with a $180$-$\mu\text{m}$-wide center conductor (aligned parallel to the $z$ axis) separated from the ground plane by two $80$-$\mu\text{m}$-wide insulating gaps. The chip is cleaned with an Ar$^+$ plasma prior to mounting under nitrogen atmosphere. Charges $Q_\text{g}$ and $Q_\text{s}$ (surface charge densities $\sigma_\text{g}$ and $\sigma_\text{s}$, respectively) can accumulate in the insulating gaps of the waveguide~\cite{Carter2012} and on an insulating layer at the surface of the superconductor. The insulating layer could originate from adsorption of residual gases at the surface during cooldown~\cite{Thiele2014} or from surface oxidation~\cite{Barends2010a}. These charges represent a potential source of stray fields $\vec{F}^\text{str}(\vec{r})$ in our experimental region.

The region above the chip where the stray electric field is measured coincides in the ($x$,$y$) plane with the cross section of the atomic beam, limited by collimating slits to the area marked by a dashed blue line in Fig.~\ref{fig:Setup}, and along $z$ by the length of the Rydberg-atom cloud of $1~\text{mm}$. Because of the translational invariance of the electrode configuration along $z$, $F^{(i)}_\text{z}(\vec{r})=0$, regardless of the potentials applied to the device, see Fig.~\ref{fig:Setup}. Moreover, $F^\text{str}_\text{z}(\vec{r})$ is constant and small enough that it does not cause shifts of the transition frequency of more than $\sim 1~\text{MHz}$, which is almost negligible compared to the shifts resulting from the $x$ and $y$ components of the stray field (see below). $F^\text{str}_\text{z}(\vec{r})$ can be determined from the Stark shift at a position where the stray field in $x$ and $y$ directions has been compensated. Consequently, only the $x$ and $y$ components of the stray-electric-field distribution need to be determined, and only three sets of measurements ($i=1,2,3$) in Eq.~(\ref{eq:Stark}) are required to determine $F^\text{str}_\text{x}(x,y)$ and $F^\text{str}_\text{y}(x,y)$.

\begin{figure*}[!tb] \centering \includegraphics[width=180mm]{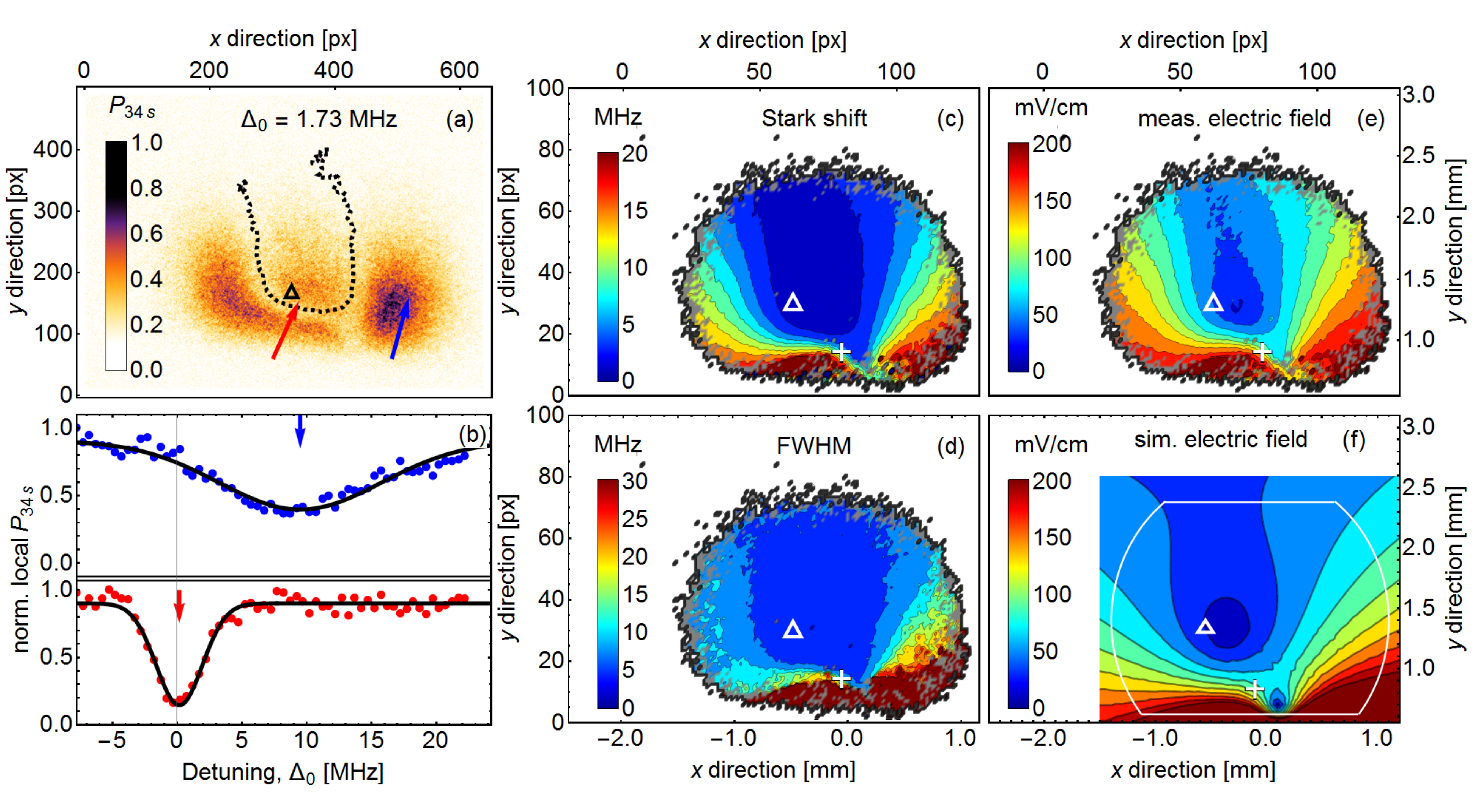} 	
\caption{Distribution of electric-field strengths for potentials $V_\text{c}=4.00~\text{V}$, $V_\text{l}=-0.02~\text{V}$, $V_\text{r}=0.00~\text{V}$ and $V_\text{s}=-0.11~\text{V}$. (a) Measured distribution of $34\text{s}$ Rydberg atoms after driving the $34\text{s}$ to $34\text{p}$ transition at a microwave detuning of $1.73~\text{MHz}$. The spatial resolution of the images is $150~\mu\text{m}$ (see panels (c)-(f) for length scales). The dots mark the positions where the field-ionization signal is minimal and represent the positions where the $34\text{s}$ to $34\text{p}$ transfer is maximal and hence the field strength (extracted from the Stark shift) is $55~\text{mV}/\text{cm}$.
(b) Experimental spectrum (dots) and Gaussian fit (solid line) of the $34\text{s}$ to $34\text{p}$ transition for He atoms located at the positions marked by the blue (upper trace) and red (lower trace) arrows in (a) after binning over $5*5$ pixels. (c) Spatial distribution of $34\text{s}$ to $34\text{p}$ transition frequencies (Stark shifts) extracted from the spectra measured at the different pixels. (d) Corresponding distribution of full widths at half maximum. (e) Electric-field distribution derived from the observed Stark shifts. (f) Simulated electric-field distribution using $\sigma_\text{g}=-23.6(1)*10^{-6}~\text{C}/\text{m}^2$ and $\sigma_\text{s}=-2.10(5)*10^{-6}~\text{C}/\text{m}^2$, see text.} \label{fig:DCFields} \end{figure*}

Propagating along the CPW, the microwave square pulses of $200~\text{ns}$ length we use to monitor the $34\text{s}$ to $34\text{p}$ transition have a Fourier-transform-limited bandwidth of $\sim4.4~\text{MHz}$ centered around a chosen detuning $\Delta_0$ from the field-free atomic resonance frequency $\nu_0$. These pulses allow us to probe and image atoms subject to specific Stark shifts and thus specific field strengths, see Fig.~\ref{fig:DCFields}(a) for an example. This is achieved by ionizing the Rydberg atoms with a pulsed field of $\sim600~\text{V}/\text{cm}$ applied $14~\mu\text{s}$ after the microwave pulse, when the Rydberg atoms have travelled more than $15~\text{mm}$ beyond the end of the chip, and extracting the electrons toward a particle detector~\cite{Thiele2014}. The detector consists of two microchannel plates in chevron configuration and a phosphor screen, imaged by a CCD camera. An einzel lens operated at field strengths up to $\sim1.5~\text{kV}/\text{cm}$ expands the electron cloud by a factor more than $6$ in $x$ and $y$ directions in a magnetic-field-free flight region of $\sim15~\text{cm}$ length, see suppl.~material for more information. Because the $34\text{p}$ state lifetime is less than $2~\mu\text{s}$, only atoms in the $34\text{s}$ state are detected. The images therefore reveal regions of low intensities [dashed line in Fig.~\ref{fig:DCFields}(a)] with a spatial resolution varying between $\leq 80~\mu\text{m}$ and $\sim150~\mu\text{m}$ which reflect the positions where atoms are efficiently transferred to the $34\text{p}$ state in the experimental region.

Well-defined electric-field distributions are generated by applying potentials $V_\text{c}, V_\text{l},V_\text{r}$ and $V_\text{s}$ to the device (Fig.~\ref{fig:Setup}). The applied potentials are small enough that they do not change the stray electric fields. The stray-field-measurement procedure outlined below starts once the stray fields have been roughly compensated by varying the potential applied to the center conductor, a potential of $4~\text{V}$ being sufficient to reduce the Stark shifts to below $25~\text{MHz}$ over the entire experimental region marked by the dashed blue line in Fig.~\ref{fig:Setup}.

We determine $\Delta E_{\text{Stark}}(x,y)/h$ by varying the detuning $\Delta_0$ of the microwave pulses between $-7~\text{MHz}$ and $22~\text{MHz}$ and extracting the intensity at every pixel of the acquired camera images. This intensity corresponds to the relative population $P_{34\text{s}}(x,y)$ in the $34\text{s}$ Rydberg state after the microwave pulse. The amplitude of the microwave pulse is small enough to avoid coherent repopulation of the $34\text{s}$ state. The spectra at each pixel are fitted by Gaussian functions to determine the width and detuning (i.e., the Stark shift) of the $34\text{s}$ to $34\text{p}$ transition. Fig.~\ref{fig:DCFields}(b) shows, as examples, the spectra recorded for $V_\text{c}=4.00~\text{V}$, $V_\text{l}=-0.02~\text{V}$, $V_\text{r}=0.00~\text{V}$ and $V_\text{s}=-0.11~\text{V}$ at the two pixels marked by red and blue arrows in the image depicted in Fig.~\ref{fig:DCFields}(a). The fits yield the values $\Delta E_{\text{Stark}}/h=0.15~\text{MHz}$ and $9.48~\text{MHz}$, respectively, with corresponding widths of $4.2~\text{MHz}$ and $14.67~\text{MHz}$. The line broadening observed with increasing Stark shift is a consequence of the field inhomogeneity and the quadratic nature of the Stark effect at low fields, as explained in Fig.~4(a) of~\cite{Osterwalder1999}.

The measured distribution of Stark shifts [Fig.~\ref{fig:DCFields}(c)] ranges from $\leq0.1~\text{MHz}$ (point marked by the label '$\Delta$') to $22~\text{MHz}$ (below the point labeled '$+$'), which is located close to the CPW. This range corresponds to electric-field strengths between $\leq16~\text{mV}/\text{cm}$ and $200~\text{mV}/\text{cm}$, see Fig.~\ref{fig:DCFields}(e). Because of the relationship between Stark shift and Stark broadening mentioned above, the observed distribution of linewidths depicted in Fig.~\ref{fig:DCFields}(d) provides an independent measurement of the stray-field distribution.

Over a large fraction of the cross section of the atomic beam, the observed microwave transitions exhibit a full-width-at-half-maximum of $\sim4.5~\text{MHz}$ [Fig.~\ref{fig:DCFields}(d)], close to the Fourier-transform limit of $\sim4.4~\text{MHz}$ of the microwave pulse. This observation implies that the effects of inhomogeneous broadening that eventually limit coherence are negligible on the $200~\text{ns}$ timescale of the measurement.

Using a $2$ dimensional finite-element calculation, we simulate the electric-field distribution~[Fig.~\ref{fig:DCFields}(f)] using the applied potentials $V_\text{c}$, $V_\text{l}$, $V_\text{r}$ and $V_\text{s}$ and fitting the surface charge densities $\sigma_\text{g}$ and $\sigma_\text{s}$ (extracted best fit values given in the caption). The simulated field distribution is in good agreement with the experimental results and the comparison indicates that $\sigma_\text{g}$ and $\sigma_\text{s}$ are defined with an accuracy of about $1\%$. Small discrepancies between the measured and simulated electric-field distributions result from the assumption that the charge density $\sigma_\text{s}$ is homogeneous over the chip surface, which is the least stringent assumption we can make on the spatial distribution of these surface charges. The simulation also confirms the conversion factor of $\sim23~\mu\text{m}/\text{px}$ relating the camera pixels to the physical dimensions in the experimental region, determined independently from simulations of the metastable He beam propagation and from simulations of the electron trajectories. From Fig.~\ref{fig:DCFields}(c-d), we extract an atom beam diameter of about $2.5~\text{mm}$.

\begin{figure*}[!tb] \centering 	
\includegraphics[width=180mm]{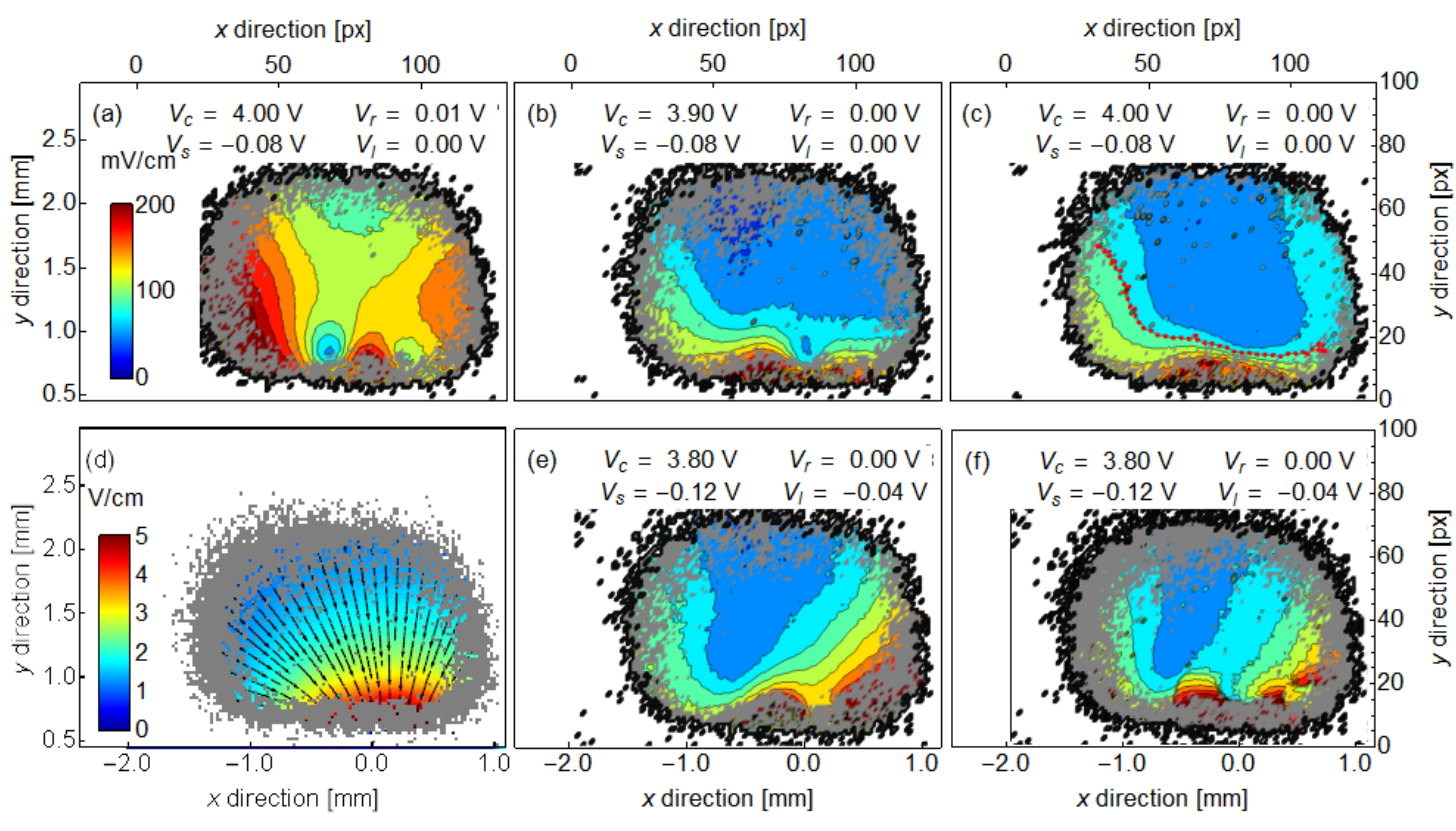} \caption{(a-c) Distribution of stray-electric-field strengths extracted from Stark shifts measured for the potential configurations defined in each panel. The dotted line in (c) indicates a region of well-compensated stray fields (see text). (d) Electric-stray-field vectors extracted from measurements (a-c). (f) Predicted electric-field distribution as measured in panel (e), see text. Except for panel (d) all color scales are described by the color legend given in (a).} 	  \label{fig:FieldComparison} 	 \end{figure*}

To determine the components of the electric stray field $(F_\text{x}^\text{str}(x,y),F_\text{y}^\text{str}(x,y))$, we measure three different electric-field distributions ($F^{(i)}_\text{tot}(x,y)$, $i=1,2,3$), for three different potential configurations [see Fig.~\ref{fig:FieldComparison}(a,b,c)]. The small homogeneous field component $F_\text{z}^\text{str}\approx50~\text{mV}/\text{cm}$ was determined from the Stark shift of $\sim1~\text{MHz}$ at the field minima in Figs.~\ref{fig:FieldComparison}(a,b,c).

$(F_\text{x}^\text{str}(x,y),F_\text{y}^\text{str}(x,y))$ are extracted from a least-squares fit of the field strength $|\vec{F}^{(i)}(x,y)+\vec{F}^\text{str}(x,y)|$ to the measured electric-field strength $F^{(i)}_\text{tot}(x,y)$ for $i=1,2,3$ using a random-search algorithm. The applied fields $\vec{F}^{(i)}(x,y)$ were obtained from finite-element calculations based on the applied voltages and the given geometry. The reconstructed stray-field vectors displayed in Fig.~\ref{fig:FieldComparison}(d) all point toward the CPW, which reveals a negative potential difference between the CPW and the metallic shield and suggests an accumulation of negative charges (electrons) near the CPW.

To test the validity of the reconstructed stray-field vectors [Fig.~\ref{fig:FieldComparison}(d)], a fourth potential configuration was explored, for which the field distribution depicted in Fig.~\ref{fig:FieldComparison}(e) was measured. Comparison of this distribution with the field distribution displayed in Fig.~\ref{fig:FieldComparison}(f) obtained by adding the measured stray field [Fig.~\ref{fig:FieldComparison}(d)] and the calculated applied field indicates excellent agreement in the field strength, with deviations of at most $50~\text{mV}/\text{cm}$ in the region where the fields could be measured.

The electric-field distribution depicted in Fig.~\ref{fig:FieldComparison}(c) corresponds to a situation where the stray electric field is almost perfectly compensated over the broad region located above the red dotted line. In this region, $\sqrt{F^{2}_\text{tot,x}(x,y)+F^{2}_\text{tot,y}(x,y)}$ does not exceed $30~\text{mV}/\text{cm}$ after compensation, which corresponds to a reduction by a factor of more than $50$ compared to the uncompensated stray-field distribution [Fig.~\ref{fig:FieldComparison}(d)].

A similar technique, based on measurements of the spatial variation of coherent population transfer, can be used to determine the distribution of the electric-field amplitude $F_\mu(\vec{r})$ of the microwave radiation generated by the CPW. Because of the translational invariance of the experimental region along $z$, the rate with which atoms are transferred from the $34\text{s}$ to the $34\text{p}$ state (Rabi rate) only varies in the $(x,y)$-plane and is given by:
\begin{equation}\label{eq:Rabirate}
\begin{aligned}
 \Omega_\text{eff}(x,y)&=&\sqrt{(2\pi\Delta(x,y))^2+(d F^\mu(x,y))^2}\\
                       &=&\sqrt{(2\pi\Delta(x,y))^2+(s(x,y)\Theta)^2}.
\end{aligned}
\end{equation}
In Eq.~(\ref{eq:Rabirate}), $\Delta(x,y)=\nu-\nu_\text{a}(x,y)$ is the detuning of the microwave frequency from the $34\text{s}$ to $34\text{p}$ resonance frequency $\nu_\text{a}(x,y)$, as determined from the stray-electric-field measurement, see Eq.~(\ref{eq:Stark}). Because of the isotropy of the $34\text{s}$ state and the small static electric fields, the $34\text{s}$ to $34\text{p}$ transition intensity (dipole moment $d\approx917 e a_0$) does not depend on the polarization of the microwave field. $F^\mu(x,y)$ is determined by measuring $\Omega_\text{eff}(x,y)$ for a well-defined microwave amplitude $\Theta$ in the CPW and using the detuning $\Delta(x,y)$, known from the field distribution [see Fig.~\ref{fig:FieldComparison}(c)]. Using Eq.~(\ref{eq:Rabirate}), the microwave-field amplitude is determined from the single parameter $s(x,y)$, which scales the microwave amplitude in the CPW to its amplitude at the location ($x,y$).

\begin{figure}[t] 	  \centering 	
\includegraphics[width=70mm]{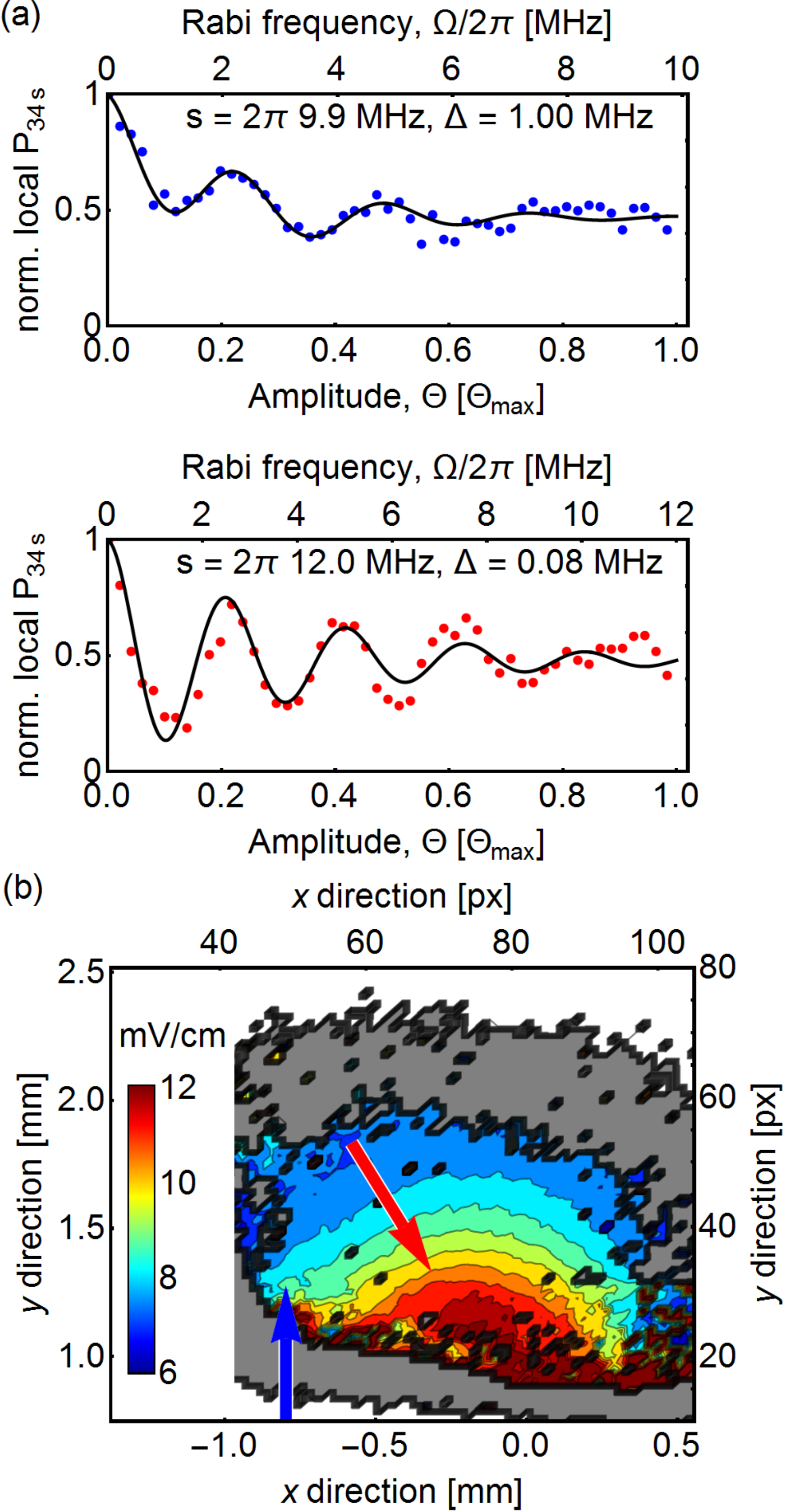} 	  \caption{(a) Coherent population transfer between the $34\text{s}$ and $34\text{p}$ state as a function of CPW drive amplitude $\Theta$ for two spatially selected atom positions marked by the blue and red arrows in (b). The black lines are fits to a simple model, see text. (b) Microwave electric field distribution for $\Theta_\text{max}$ extracted from the measured distribution of $\Omega_\text{eff}(x,y)$ using Eq.~(\ref{eq:Rabirate}). The CPW is located at the origin $(0,0)$.} \label{fig:MWFieldsMain} 	 \end{figure}

Fig.~\ref{fig:MWFieldsMain}(a) presents two measurements of the $34\text{s}$ to $34\text{p}$ population transfer carried out at the positions marked with arrows in Fig.~\ref{fig:MWFieldsMain}(b) by varying $\Theta$ from $0~\text{V}$ to $\Theta_\text{max}\approx 560~\mu\text{V}$ in the center conductor and using a $400$-$\text{ns}$-long microwave pulse of frequency $27.966~\text{GHz}$.
Because the stray-field distribution [Fig.~\ref{fig:FieldComparison}(c)], the displacement of the $34\text{s}$ to $34\text{p}$ transition, and the lifetime of the $34\text{p}$ state are known, $s(x,y)$ can be unambiguously determined using a simple model (black lines) based on Rabi's formula (see Eq.~(1) in \cite{Thiele2014}). The results are depicted in Fig.~\ref{fig:MWFieldsMain}(b). The large dipole moment of the $34\text{s}$ to $34\text{p}$ transition implies that a microwave field amplitude as weak as $\sim1.0~\text{mV}/\text{cm}$ suffices to fully transfer the population within $400~\text{ns}$.

Overall, the measured microwave field decreases with increasing distance from the CPW, as expected. However, the observed field maximum of the mode does not exactly lie above the CPW, but $\sim250~\mu\text{m}$ to its left. Motivated by numerical simulations, this shift could be explained by interfering even and odd modes of the CPW because of different ground potentials of the (super)conducting surfaces with lateral dimensions larger than the microwave wavelength.

In conclusion, we have presented a technique for measuring distributions of electric-field vectors with a spatial resolution of about $100~\mu\text{m}$ exploiting spatial variations of Rydberg Stark shifts and coherent population transfer in a supersonic beam of helium Rydberg atoms propagating above a cryogenic chip. By measuring several distributions of Stark shifts for the $34\text{s}$ to $34\text{p}$ transition resulting from different sets of potentials applied to the chip electrodes, the measured stray field could be accurately determined. The stray-field vectors allowed the identification of its source - in our case negative charges accumulating on the insulating surfaces in the vicinity of the CPW - and the determination of optimal potential configurations for stray-field compensation.

Exploiting coherent population transfer between the $34\text{s}$ and $34\text{p}$ Rydberg state, a similar technique allowed the measurement of the distribution of microwave amplitudes above the CPW. The reconstructed distribution of microwave field amplitudes did not exactly coincide with the expected mode of a CPW but revealed the effects of other interfering modes. By adding more microwave sources on the chip surface, it should be possible to fully characterize these spurious modes with the same technique as used to determine the stray electric-vector field.

For experimental setups that are not translationally invariant, it is straightforward to characterize the electric field along the beam propagation axis ($z$ direction). For different start times of the microwave pulse that are applied to the coplanar waveguide, the finite velocity of the atoms in $z$ direction causes the atom cloud to be localized at different positions along the beam propagation axis at which electric fields can be measured~\cite{Thiele2014}. Also, the methods we presented are not limited to experiments with atomic beams but also applicable, e.g., to experiments using ensembles of ultracold atoms.

The techniques presented in this article have several advantages compared to previous works using (Rydberg) atoms to determine electric fields. The data can be acquired rapidly, because only a single spectrum is enough to measure a full $2$-dimensional electric field distributions, and a minimal set of $4$ spectra characterizes all stray fields. Electric-field distributions are determined on a mm-sized area, limited only by the cross-section of the atomic beam. The spatial resolution of the Rydberg-atom images above the surface is currently limited to $\sim 100~\mu\text{m}$ in both transverse dimensions by the transverse velocity spread of the Rydberg atom beam. Simulations of the electron-trajectories through optimized electron optics in combination with pulsed-field ionization indicate that laser cooling in the transverse dimension to the Doppler limit of $1\text{s}^12\text{s}^1~{}^{3}S_{1}$ helium atoms ($0.28~\text{m}/\text{s}$) should enable an improvement of the resolution to $4~\mu\text{m}$. In contrast to other reported techniques, it does not alter the properties of the surface as it only requires low-frequency radiation (no laser radiation) close to the surface, and because of the use of an inert gas that does not adsorb to the surface. Finally, it is compatible with cryogenic temperatures.

Additionally, the spatial selection of atoms in field-free regions paves the way for precision measurements of atoms close to surfaces useful to study Van-der-Waals interactions~\cite{Anderson1988} or surface ionization~\cite{Nordlander1955,Neufeld2008,Pu2013}. Finally, the methods represent an important step toward coupling single Rydberg atoms to microwave photons on a chip.

\textbf{Acknowledgements}:
We thank Stefan Filipp for his contributions to the initial phase of the experiment, Valentin Goblot and Florian L\"uthi for the setup and characterization of the active magnetic shielding, and Pitt Allmendinger for useful discussions.

We acknowledge the European Union H$2020$ FET Proactive project RySQ (grant N. $640378$). Additional support was provided by the Swiss National Science Foundation (SNSF) under project number $20020\_149216$ (FM) and by the National Centre of Competence in Research "Quantum Science and Technology" (NCCR QSIT), a research instrument of the SNSF.

\end{document}